\documentclass[aps,twocolumn,showpacs,preprintnumbers,amsmath,amssymb,prl]{revtex4}
\usepackage{natbib}
\usepackage{amsmath}
\usepackage{amssymb}
\usepackage{graphicx}
\usepackage{color}

\newcommand{\mx}{\mbox}

\begin{document}
\setlength{\unitlength}{1mm}

\title{Self-Calibration of CMB Polarization Experiments}
\author{
Brian G. Keating$^{1}$, Meir Shimon$^{2}$, and Amit P.S. Yadav$^{1}$}
\affiliation{
$^1$Center for Astrophysics and Space Sciences, Department of Physics, University
of California, San Diego, 9500 Gilman Drive, La Jolla, CA, 92093-0424\\
$^2$School of Physics and Astronomy, Tel Aviv University, Tel Aviv 69978, Israel\\
}
\date{\today}%

\begin{abstract} 
Precision measurements of the polarization of the cosmic microwave 
background (CMB) radiation, especially experiments seeking to detect the odd-parity ``B-modes'', 
have far-reaching implications for cosmology.   To detect the B-modes generated during inflation the flux response and polarization angle of these experiments must be calibrated to exquisite precision. While suitable flux calibration sources abound, polarization angle calibrators are deficient in many respects. Man-made polarized sources are often 
not located in the antenna's far-field, have spectral properties that are radically different from the CMB's, 
are cumbersome to implement and may be inherently unstable over the (long) duration these searches require to 
detect the faint signature of the inflationary epoch. Astrophysical sources suffer from 
time, frequency and spatial variability, are not visible from all CMB observatories, and none are understood with 
sufficient accuracy to calibrate future CMB polarimeters seeking to probe inflationary energy scales of 
$10^{15}$ GeV.  Furthermore, both man-made and astrophysical sources are often much brighter than the CMB B-mode signal and these bright sources can cause non-linearities in the detector's response. Both  man-made and astrophysical sources require dedicated observations which
detract from the amount of integration time usable for detection of the inflationary B-modes. CMB $TB$ and $EB$ modes, expected to identically vanish 
in the standard cosmological model, can be used to calibrate CMB polarimeters. By enforcing the observed $EB$ and $TB$ 
power spectra to be consistent with zero, CMB polarimeters can be calibrated to levels not possible 
with man-made or astrophysical sources. All of this can be accomplished without any loss of observing time using a calibration source which is spectrally identical to the CMB B-modes. The calibration procedure outlined here can be used for any CMB polarimeter.
%irrespective of the science goal assuming no cosmic-birefringence, i.e. standard cosmological model is parity symmetrical.
\end{abstract}

\maketitle
                                                                                
\mx{\it {Introduction}}: 
Inflation is perhaps the most promising model of the early universe, resolving the Big Bang model's flatness and horizon problems and providing seed 
perturbations for structure formation (see \emph{e.g.,} ~\cite{baumann_cmbpol} for review). 
Besides the density, or ``scalar'', seed perturbations, inflationary cosmological models also predict
``tensor'' perturbations arising from a primordial gravitational wave background.  Primordial scalar
perturbations create only CMB $E$-modes,
%~\footnote{To first order in perturbations, primordial scalar perturbations do not generate CMB B-modes.  However at second (and higher) order in the perturbations, scalar perturbations do produce B-modes~\cite{Bartolo:2006fj,Baumann:2007zm}.The B-modes generated from higher order perturbations are expected to be smaller than the tensor $B$-mode levels that the upcoming and future experiments (like CMBPol) are sensitive to.}, 
while primordial tensor perturbations generate both parity-even $E$-modes 
and parity-odd $B$-modes polarization~\cite{SeljakZaldarriaga97,1997PhRvD..55.7368K,1997PhRvL..78.2058K}. 
The detection of primordial tensor $B$-modes in the CMB would confirm 
the existence of gravitational wave perturbations in the early universe.  
Numerous observational efforts are underway to detect the CMB's $B$-mode since such a detection would establish the 
energy scale at which inflation occurred. The amplitude of primordial $B$-modes can be characterized by the tensor-to-scalar ratio, $r$. The most restrictive limit on $r$ is currently $r < 0.18$ ($95\%$ confidence)~\cite{Story2012}, and the best direct limits on $r$ from $B$-mode measurements is $r<0.72$ ($95\%$ confidence)~\cite{Chiang_et_al_2010}.

An impressive battery of CMB experiments have mapped the CMB's intensity to near
cosmic-variance-limited precision to $\ell\sim3000$\cite{ACT, Story2012, wmap7_cosmology}. 
However, the measurement of the CMB's $E$-mode polarization signal is considerably more challenging since it is 10 to 100 times smaller than the CMB's temperature anisotropy. Compared to these signals the $B$-mode polarization from inflationary gravitational waves is even more challenging -- current upper limits correspond to $B$-mode fluctuations less than $10\%$ of the E-modes. Calibration accuracy required by a given instrument to constrain or detect the minute $B$-mode signal is well-discussed in the literature~\cite{HHZ}.
Calibration is accomplished either using hardware calibrators, located in the near field of 
the instrument~\cite{Keating2003, Takahashi2010, Hinderks2009, Tajima2012}, or by measurements 
of polarized astrophysical sources, \emph{e.g.} the Moon, Tau A, Cen A, 3C 273, or the galactic 
plane \cite{Aumont, Zemcov, Agudo2012,2010SPIE.7741E..77M}. Neither hardware polarization 
calibrators nor astrophysical sources can achieve better than $\simeq 1^\circ$ precision on the polarization angle calibration. 
It is hard, if not impossible, to do better, yet a precision of $1^\circ$ is insufficient for detecting $r=0.025$ to $0.01$~\cite{HHZ} -- the goal of future polarimeters. Rather, it has been shown~\cite{Miller2009} that for $r=0.01$ to be biased by $\lesssim 0.1\sigma_{r}$ (where $\sigma_{r}$ is the nominal 
statistical uncertainty on the inferred value of $r$), the uncertainty in pixel rotation cannot exceed $4'$ even for a nearly ideal CMB experiment.
Even relaxing this requirement, by allowing a bias on $r$ as large as $\simeq 1\sigma_{r}$, only permits a
miscalibration level of $\lesssim 12'$. 

%\textcolor{red}{see my email... i think it's quick to just do the sum and then it's independent from experiment to experiment. %in fact, as you point out, it's really a  lower limit which is fine for this paper}

Miscalibration of the instrument's polarization angle (pixel rotation)
mixes polarization modes, leaking $E$- into $B$-modes, thereby
producing spurious B-mode polarization~\cite{HHZ,Shimon2007,YSZ09}.  
Additionally, due to polarization mode-mixing, new $TB$ and 
$EB$ correlations are generated.  Since the standard cosmological model is 
parity-even the $TB$ and $EB$ correlations identically
vanish. Therefore the $TB$ and $EB$ spectra can be used to probe the miscalibration of the pixel rotation angle. 
Furthermore, the miscalibration angle \emph{itself} can be quantified, then removed, resulting in an unbiased measurement of B-mode polarization. This calibration procedure is accomplished \emph{during} data acquisition, requiring no additional observing time. Therefore it is referred to as ``self-calibration". Moreover, since the calibration signal 
is the CMB itself, any concerns that the detector response will behave nonlinearly are eliminated.    

It has been shown that miscalibration produces 
a distinct signature in the (otherwise zero) $\langle EB\rangle$ 
and $\langle TB \rangle$ correlations. The amplitude of these ``forbidden'' correlations is proportional to the amount of miscalibration and, furthermore, it is known that several other instrumental systematics can be detected using these $EB$ 
and $TB$ correlations ~\cite{Miller2009, YSZ09}.  This paper uses these correlations to calibrate CMB polarimeters 
to levels not achievable with laboratory or astrophysical sources. 

%Let us emphasize that calibrating polarization angles has been a significant 
%challenge in the CMB field.  
%In this paper we show that we do not need such radical step. 
%For this reason we only focus on calibrating the polarization angle.

\mx{\it {Polarization Map Making and Miscalibration}}: 
Following~\cite{Chiang_et_al_2010}, the timestream data from a 
single polarization sensitive detector, $d_{i}$, is written as
\begin{equation}
d_{i} = g_{i} \left[T({\bf p}) + \gamma_{i} (Q({\bf p})\cos2\psi_{i} + 
U({\bf p})\sin2\psi_{i})\right],
\label{eq:psb_timestream}
\end{equation}
where $g_i$ is the flux calibration, or ``gain'' for the $i$-th detector, $T, Q, U$ are the beam-integrated CMB Stokes
parameters for the map pixel in direction ${\bf p}$, $\gamma_i \equiv (1 - \epsilon_i) / (1 +
\epsilon_i)$ is the polarization efficiency factor, $\epsilon_i$ is the polarization leakage for the $i$-th detector, and $\psi_i$ is the
detector's polarization orientation projected on the sky. The goal of
mapmaking is to recover $T, Q, U$ from the detector timestreams.

The angle $\psi_i$ is modeled as 
\begin{eqnarray}
\psi_i=\psi_{\text{design}}+\Delta\psi\,,
\label{deltapsi}
\end{eqnarray} 
where $\psi_{\text{design}}$ is the intended orientation of the detector on the sky with respect to right ascension and declination and $\Delta\psi$ is the miscalibration of the detector. Figure 1 displays the coordinate system used for a single polarization sensitive detector.

\begin{figure}
\includegraphics[height=0.44\textwidth,angle=0]{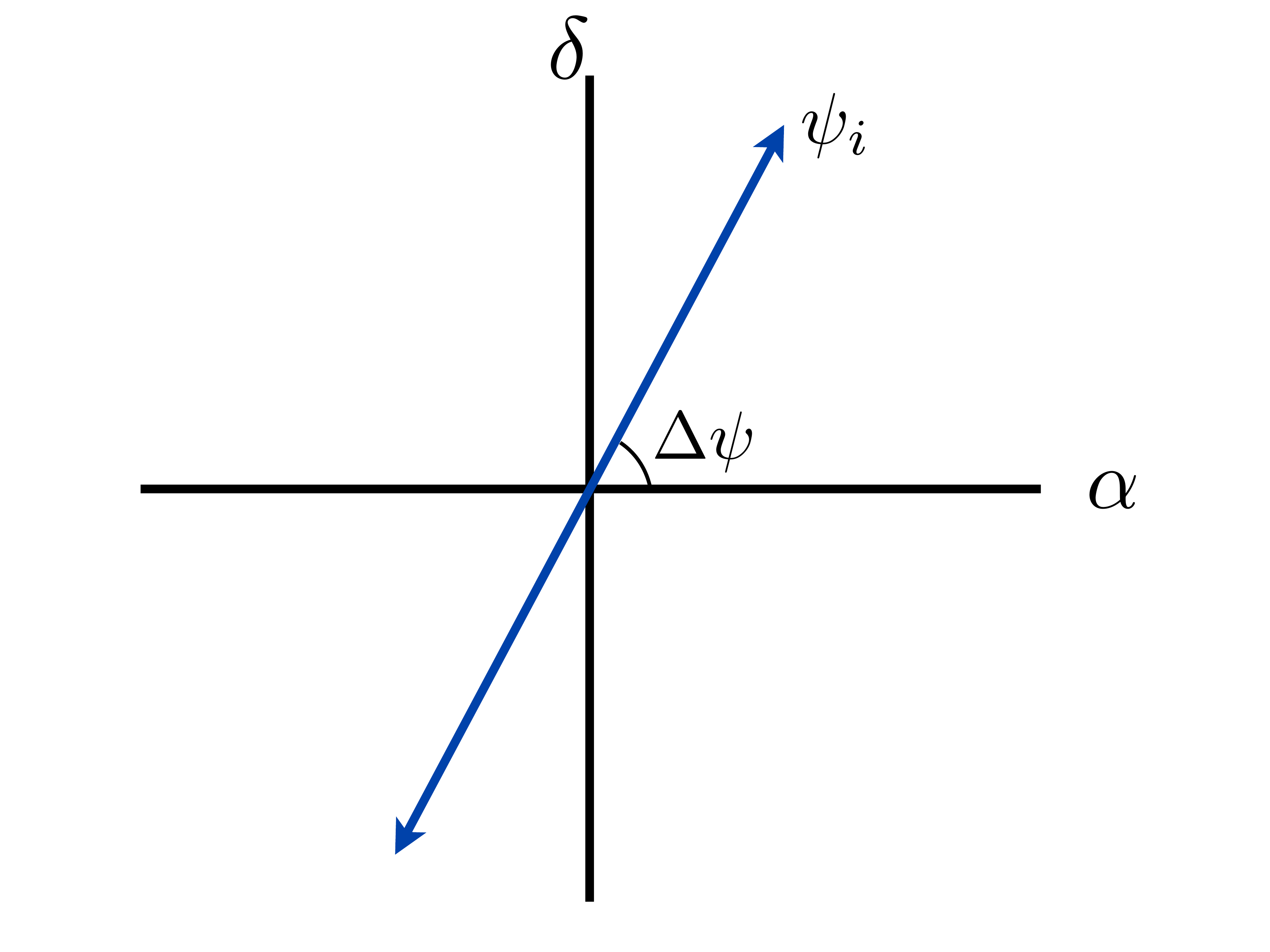}
\caption{Coordinate system showing the relevant angles from equation (2) for a single polarization sensitive detector. The horizontal and vertical axes are right accession ($\alpha$) and declination ($\delta$). For this detector $\psi_{\text{design}}$ was intended to be parallel to $\alpha$.}
\label{fig1}
\end{figure}

\begin{figure*}
 \includegraphics[height=0.60\textwidth,angle=0]{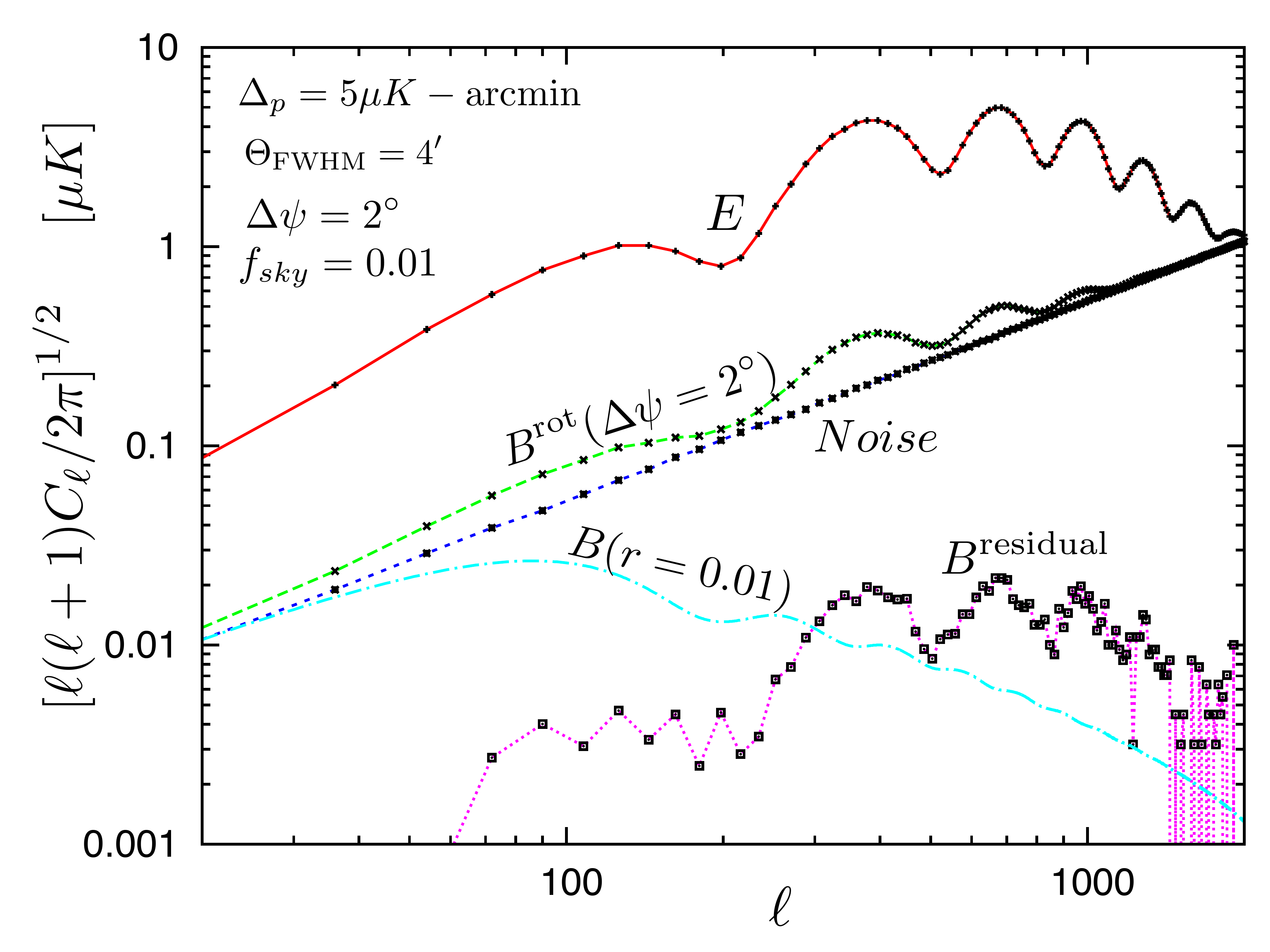}
  \caption{Simulation demonstrating the self-calibration of pixel rotation using  
the CMB's $EB$ power spectrum. The polarization angle's miscalibration angle is assumed to 
be $\Delta\psi=2^{\circ}$, and the instrumental noise is $\Delta_E=\Delta_B=5 \mu K$-arcmin. The
beam's full-width at half-maximum is $\Theta_{\text{FWHM}}=4'$.  
The solid red curve shows the $E$-mode power spectrum, the green long-dashed curve 
shows the $B$-modes induced by miscalibration. The blue short-dashed curve shows the $B$-mode after correcting for the miscalibrated pixel rotation angle. 
Finally, to demonstrate how well the self-calibration works, the instrumental noise contribution is subtracted from the de-rotated $B$-mode in
the bottom-most (magenta dotted) curve; these are residual B-modes remaining after self-calibrating the instrument. The primordial B-mode spectrum corresponding to $r=0.01$ is shown in dot-dashed (cyan) line. For clarity, the lensing $B$-modes are omitted.}
  \label{fig2}
\end{figure*}

Calibration of the detector's pixel rotation is one of the most challenging 
tasks facing the experimentalists~\cite{Takahashi2010, QUIET_instrument}. A miscalibration of the detector's
polarization angle by an amount $\Delta\psi$ rotates 
primordial Stokes parameters $\tilde Q({\bf n})$, and $\tilde U({\bf n})$ into the observed quantities:
\begin{eqnarray}
Q ({\bf n})\pm iU({\bf n})=e^{\pm 2i \Delta\psi}(\tilde Q({\bf n})\pm i \tilde U({\bf n}))\,.
\end{eqnarray}
The measured Fourier modes, $E({\bf l})$ and $B({\bf l})$, written in terms 
of the primordial modes $\tilde E({\bf l})$ and $\tilde B({\bf l})$, become
\begin{eqnarray}
E({\bf l})&=&\cos{(2\Delta\psi)} \tilde E({\bf l}) + \sin{(2\Delta\psi)} \tilde B({\bf l}) \nonumber \\
B({\bf l})&=&-\sin{(2\Delta\psi)} \tilde E({\bf l}) + \cos{(2\Delta\psi)} \tilde B({\bf l})\,.
\end{eqnarray}

The above equations show that pixel rotation modifies the 
power spectra of $E({\bf l})$ and $B({\bf l})$ and generates 
spurious correlation between $E({\bf l})$ and $B({\bf l})$ and 
between $T({\bf l})$ and $B({\bf l})$, modifying the observed power spectra as follows:
\begin{eqnarray}
C^{TE}_\ell  &= &\cos{(2\Delta\psi)} \tilde {C}^{TE}_\ell \nonumber \\
C^{EE}_\ell&=&\sin^2{(2\Delta\psi)} \tilde C^{BB}_{\ell} + \cos^2{(2\Delta\psi)} \tilde C^{EE}_{\ell} \nonumber \\
C^{EB}_\ell&=&\frac{1}{2}\sin{(4\Delta\psi)} (\tilde C^{BB}_\ell - \tilde C^{EE}_{\ell}) \nonumber \\
C^{TB}_\ell&=&-\sin{(2\Delta\psi)} \tilde C^{TE}_\ell \nonumber \\
C^{BB}_\ell&=&\cos^2{(2\Delta\psi)} \tilde C^{BB}_\ell + \sin^2{(2\Delta\psi)} \tilde C^{EE}_\ell\,.
\label{eq:spectra}
\end{eqnarray}
Here, and throughout, tildes represent primordial quantities. From Eq. \ref{eq:spectra} it is clear that pixel rotation generates spurious $B$-modes in the absence of primordial B-modes. Calibration of the pixel rotation involves finding $\Delta\psi$ for the detector system and removing it from the data prior to map making (Eq. \ref{eq:psb_timestream}). This procedure will recover the unrotated, primordial CMB polarization spectra.

In the absence of cosmological parity violation leading to cosmic birefringence \cite{1999PhRvL..83.1506L, Alexander},
the detection of $EB$ and $TB$ spectra -- each having the \emph{same} implied pixel rotation 
angle $\Delta\psi$ --  directly implies that the detector's polarization angles have been
miscalibrated. Such miscalibration can be caused, \emph{e.g.}, by fabrication errors or sources of birefringence in the telescope's optics.
%~\footnote{If the cosmic birefringence is spatially varying \emph{e.g.}, 
%\cite{2012arXiv1206.5546G}, $EB$ and $TB$ will produce \emph{different} implied $\Delta\psi$ values. 
%Only when the TB and EB estimators agree }.The significance of the detection increases as the instrumental systematics are measured better through self-calibration. 

%\emph{YES! THIS IS WHY WE NEED TO SHOW HOW WE MEASURE $\psi$!}
%
%\textcolor{blue}{I've addressed this second point in the note I just sent you, just 
%plug in the numbers in Eq.(10) of the note.}

The miscalibration angle $\Delta\psi$, is obtained from the observed $C^{TB}_\ell$ and $C^{EB}_\ell$ by minimizing the variance of the 
difference between the observed and theoretical power power spectra as a function of $\Delta\psi$.

Using Eq. \ref{eq:spectra}, the two independent likelihood functions for the miscalibrated pixel rotation angle, $\Delta\psi$, become

\begin{eqnarray}
\mathcal{L}_{TB}&\propto &\exp\left[-\sum_{l}\frac{(C_{l}^{TB}+\sin 2\Delta\psi\tilde{C}_{l}^{TE})^{2}}{2(\delta C_{l}^{TB})^{2}}\right]\nonumber\\
\mathcal{L}_{EB}&\propto &\exp\left[-\sum_{l}\frac{(C_{l}^{EB}+\frac{1}{2}\sin 4\Delta\psi\tilde{C}_{l}^{EE})^{2}}{2(\delta C_{l}^{EB})^{2}}\right]
\label{eq:likes}.
\end{eqnarray}

For simplicity, in Eq. \ref{eq:likes} it is assumed that $\tilde{C}_{l}^{EE} \gg \tilde{C}_{l}^{BB}$, and that

\begin{eqnarray}
(\delta C_{\ell}^{TB})^{2}&=&\frac{1}{(2\ell+1)f_{sky}}C_{\ell}^{TT,tot}C_{\ell}^{BB,tot}\nonumber\\
(\delta C_{\ell}^{EB})^{2}&=&\frac{1}{(2\ell+1)f_{sky}}C_{\ell}^{EE,tot}C_{\ell}^{BB,tot}
\end{eqnarray}
where $f_{sky}$ is the fraction of the sky observed, $X,Y\in\{T,E,B\}$, and

\begin{eqnarray}
 C_\ell^{XY,{\text tot}}=\tilde C_\ell^{XY}+\delta^{XY}\Delta^2_X e^{\ell^{2}\Theta^2_{\text{FWHM}}/(8\ln{2})}.
\end{eqnarray} Here, $\Delta_X$ is the detector noise and  $\Theta_{\text{FWHM}}$ is the 
full-width at half-maximum (FWHM) resolution of the polarimeter's Gaussian beam. 

The best-fit estimates for the pixel rotation angle are obtained by maximizing the likelihood functions, Eq. \ref{eq:likes}, resulting in
%$C^{TB}_\ell$ and $C^{EB}_\ell$ respectively, are
\begin{eqnarray}
\Delta\psi_{TB}&=&\textcolor{blue}{-}\frac{1}{2}\sin^{-1}\Bigg(\frac{A_{TB}}{B_{TB}}\Bigg)\nonumber\\
\Delta\psi_{EB}&=&\textcolor{blue}{-}\frac{1}{4}\sin^{-1}\Bigg(\frac{2A_{EB}}{B_{EB}}\Bigg)
\label{eq:psis}
\end{eqnarray}

where \begin{eqnarray}
A_{TB}&=&\sum_{l}\frac{2\ell +1}{2}\frac{C^{TB}_\ell C^{TE}_\ell} {C_\ell^{TT,tot} C_\ell^{BB,tot}}\nonumber\\
B_{TB}&=&\sum_{l}\frac{2\ell +1}{2}\frac{(C^{TE}_\ell)^{2}} {C_\ell^{TT,tot} C_\ell^{BB,tot}}\nonumber\\
A_{EB}&=&\sum_{l}\frac{2\ell +1}{2}\frac{C^{EB}_\ell C^{EE}_\ell}{C_\ell^{EE,tot}C_\ell^{BB,tot}}\nonumber\\
B_{EB}&=&\sum_{l}\frac{2\ell +1}{2}\frac{(C^{EE}_\ell)^{2}} {C_\ell^{EE,tot}C_\ell^{BB,tot}},
\end{eqnarray}
and $X\in\{T,E,B\}$. 

The values for $\Delta\psi_{EB}$ and $\Delta\psi_{TB}$ obtained from observations, Eqns. \ref{eq:psis}, should agree to within the statistical uncertainty (which is derived next). The consistency $\Delta\psi_{TB}\simeq \Delta\psi_{EB}$ provides a powerful cross-check on the hypothesis that the pixel rotation has been miscalibrated. Since there is intrinsic correlation between $EB$ and $TB$, for real data Eqns. \ref{eq:likes} should be replaced with a single likelihood function that {\it simultaneously} uses $C^{TB}$, $C^{EB}$ and their correlation, to infer a single $\Delta\psi$ value. In practice, $\Delta\psi_{EB}$ will be superior to $\Delta\psi_{TB}$ but the value of having two independent estimates, which must agree in order to apply the self-calibration method, motivates the construction of both estimators.

The signal-to-noise for the $EB$ and $TB$ detection is
\begin{eqnarray}
\left(\frac{S}{N}\right)_{TB}^{2} &=&f_{sky}\sum\frac{2\ell+1}{2}
\frac{(C_\ell^{TB})^{2}}{{C}_\ell^{TT,tot}{C}_\ell^{BB,tot}}\nonumber\\
\left(\frac{S}{N}\right)_{EB}^{2}&=&f_{sky}\sum\frac{2\ell +1}{2}
\frac{(C_\ell^{EB})^{2}}{{C}_\ell^{EE,tot}{C}_\ell^{BB,tot}}
\end{eqnarray}
where $f_{sky}$ is the fraction of the sky observed.
% and $C_{l}^{XX}$ are the observed power spectra
%including both the signal and the instrumental noise (here $X\in\{T,E,B\}$)
% \begin{eqnarray}
%\tilde C_\ell^{XX,{\text tot}}=\tilde C_\ell^{XX}+\Delta^2_X e^{\ell^{2}\theta^2_{\text{FWHM}}/(8\ln{2})}\,,
%\end{eqnarray}
%here $\Delta_X$ is the detector noise and  $\theta_{\text{FWHM}}$ is the 
%full-width half-maximum (FWHM) resolution 
%of the polarimeter's Gaussian beam. 
The uncertainties in $\Delta\psi_{EB}$ and $\Delta\psi_{TB}$ are
\begin{eqnarray}
\sigma_{\Delta\psi_{TB}}^{2}=\left[4f_{sky}\sum\frac{2\ell +1}{2}\frac{(\tilde{C}_\ell^{TE})^{2}}{\tilde{C}_\ell^{TT,tot}\tilde{C}_\ell^{BB,tot}}\right]^{-1}\nonumber\\
\sigma_{\Delta\psi_{EB}}^{2}=\left[4f_{sky}\sum\frac{2\ell +1}{2}\frac{(\tilde{C}_\ell^{EE}-\tilde{C}_\ell^{BB})^{2}}{\tilde{C}_\ell^{EE,tot}\tilde{C}_\ell^{BB,tot}}\right]^{-1}\,.
\label{eq:psinoise}
\end{eqnarray}
  
Equations \ref{eq:psis} and \ref{eq:psinoise} give the miscalibration of pixel rotation and its uncertainty. 
Armed with these quantities, the experimentalist can go back and correct the assumed polarization angle by subtracting $\Delta\psi$ from $\psi_{i}$ in Eq. \ref{eq:psb_timestream}. For a toy experiment with $\Delta_p=5 \mu K$-arcmin, $f_{sky}=0.01$, and $\Theta_{FWHM}=4'$ a pixel rotation as small as $\Delta\psi \sim 0.05^\circ (3^\prime)$  can be detected using the $EB$ power spectrum alone, and $\Delta\psi \sim  0.1^\circ$, using the $TB$ power spectrum alone. These values are more than sufficient to detect $r=0.01$ with $\lesssim 0.1\sigma_{r}$  bias. Fig.~\ref{fig2} shows the pixel rotation calibration method derived using the $EB$ spectrum. A similar result can be obtained for $\Delta\psi_{TB}$.

\mx{\it {Discussion}}: While pixel rotation is potentially the most pernicious
obstacle to detecting primordial B-modes,  there are other systematic effects such as 
differential ellipticity and differential pointing that can produce  $TB$ and $EB$ correlations \cite{Miller2009, YSZ09}. However the $TB$ and $EB$ correlations induced by these systematic effects have different angular dependencies from pixel rotation~\cite{yadav2012}. Therefore, high-sensitivity observations covering large numbers of multipoles can make detailed measurements of the $TB$ and $EB$ correlations to quantify, and correct for these types of systematic errors. Furthermore, due to the dependence on the beamsize, for a fixed level of instrument noise, a higher resolution experiment will always
calibrate pixel rotation more precisely than a lower resolution experiment. %A joint study of pixel rotation calibration in the presence of other systematics is underway \cite{KSZ12}.

Pixel rotation and cosmological birefringence 
are fully degenerate effects; \emph{i.e.} the angle $\Delta \psi$ in Eq.~(\ref{deltapsi}) \emph{could be} the sum 
of the two effects. However, for the purpose of B-mode detection it does not 
matter what causes the polarization rotation $\Delta\psi$; we simply de-rotate the 
polarization by the angle $\Delta\psi$ inferred from the $EB$ and $TB$ estimators. Furthermore cosmic birefringence detection is not sacrificed when cosmic birefringence is spatially varying~\cite{Kamionkowski_08,Yadav_etal_09,2009PhRvD..80b3510G,YSZ09,2012arXiv1206.5546G,yadav2012}.

Complications arising from the $E-B$ separation due to partial sky coverage and the $EB$ and $TB$ correlations induced by gravitational lensing of the CMB by large scale structure are irrelevant to the self-calibration procedure proposed here because these EB and TB correlations couple only different $\ell$-values, \emph{i.e.,} $\langle E_{\ell}B_{\ell'}\rangle$ and $\langle T_{\ell}B_{\ell'}\rangle$ will be non-vanishing only when $\ell\neq\ell'$. 

%Complications arising from the $E-B$ separation due to partial sky coverage have been neglected here. However, the $EB$ and $TB$ correlations induced due to boundary effects will fluctuate around zero as a function of $\ell$ since the boundary does not define a preferred direction in the sky. Hence, this effect does not bias our estimate for the pixel rotation angle; it only affects its variance.  Furthermore, given the sky mask applied to the data, such boundary effects can be removed via Monte Carlo simulations.Finally, it is noteworthy to mention that while gravitational lensing of the CMB by large scale structure also generates $EB$ and $TB$ correlations, the modes correlated via lensing and via pixel rotations are ``orthogonal" and hence lensing does not affect the estimation of pixel rotation \cite{Yadav_etal_09,2009PhRvD..80b3510G} . 

\mx{\it {Acknowledgments}}: MS' work was supported by the UCSD-Tel Aviv University Cosmology Program. We thank Bryan Steinbach, Chang Feng, Zigmund Kermish, and  Kam Arnold for useful comments.

\bibliography{references}
%\end{thebibliography}
                                                                                
\end{document}